\documentstyle[12pt]{article}
\begin{document}
\title{\vspace{-4cm}Comment about UV regularization 
of basic commutators in string theories}
\author{A. Yu. Kamenshchik$^{1,2}$, I. M. Khalatnikov $^{1,2,3}$ and \
 M. Martellini$^{2,4}$}
\date{}
\maketitle
\hspace{-6mm}
$^{1}${\em L. D. Landau Institute for Theoretical Physics,
Russian Academy of Sciences, Kosygin Street 2, Moscow, 117334,
Russia}\\
$^{2}${\em
Landau Network-Centro Volta, Villa Olmo, Via Cantoni 1, 
22100 Como, Italy}\\
$^{3}${\em Tel Aviv University, Raymond and Sackler Faculty of Exact Sciences;
School of Physics and Astronomy, Ramat Aviv, 69978, Israel}\\
$^{4}${\em Dipartamento di Fizica dell' Universita di Milano and INFN,
Sezione di Milano, I-20133, Milan, Italy}

\begin{abstract}
Recently proposed by Hwang, Marnelius and Saltsidis 
zeta regularization of basic commutators in string theories
is generalized  to the string models with non-trivial vacuums. It is shown that
implementation of this regularization implies the cancellation of dangerous 
terms in the commutators between Virasoro generators, which break Jacobi
identity. 
\end{abstract}

It is well known what the role plays the problem of regularization and
renormalization
of ultraviolet divergences in quantum field theory. However, the importance
of the problem of the proper treatment of ultraviolet divergences arising at 
quantization of string theories is less understood. Perhaps, it is connected
with the
fact that calculating the central extensions in Virasoro algebra in string
theories
it is possible to escape the necessity of consideration of infinite sums due to 
tacitly assumed and very natural regularization providing their cancellation
\cite{string}.
However, in the cases when conventional vacuum state is absent (for example
in the case
of tensionless conformal string \cite{tensionless})  explicit ultraviolet
regularization of basic commutators of the theory becomes inevitable. 

In recent paper by Hwang, Marnelius and Saltsidis \cite{HMS} the conveninent
version
of ultraviolet zeta regularization of basic commutators in string theories was
elaborated. This scheme was applied to Hamiltonian BRST quantization \cite{BRST}
of string theories. The well-known results for critical dimensions
obtained by the same method earlier \cite{Hwang} were reproduced in the new
context.

In this short comment we further investigate the properties of
regularization suggested
in \cite{HMS} constructing two simple examples. We consider traditional
closed bosonic
string model but with non-trivial vacuums. These vacuums are obtained by
means of
redifinition of creation and annihilation operators due to Bogoliubov
transformation.
As a result of this redefinition in the expression for central extension
arise infinite
sums while the traditional finite part of contribution to the central charge
acquires
the structures breaking Jacobi identities. The direct generalization of the
regularization
suggested in \cite{HMS} allows calculate the finite contributions of
infinite sums,
which  have the structure providing the total cancellation of patalogical
terms and restoration of Jacobi identities. While our models hardly can be
treated
as having direct physical significance, the very fact of restoration of 
the correct structure of central extension looks promising. It could be
interpreted
as an additional argument in favour of regularization \cite{HMS}. One can
hope also
the this scheme can be applied to more complicated theories such as quantum 
cosmology where the necessity of ultraviolet regularization is getting
obvious \cite{semen}. 
 
The structure of the present note is the following:
first, we introduce the necessary notations for closed bosonic string and write 
down the expression for the central charge arising in the commutator between 
Virasoro generators for the family of vacuums, parametrized by form of
relation between
energy and momentum of excitations described by the corresponding creation and 
annihilation operators; second, for particular choices of vacuums we show
the possible
sources of the break of Jacobi identities; third we write down formulae for
the regularized
commutators \cite{HMS} and show how they modify  expressions for the central
extension
of Virasoro algebra; finally using zeta regularization \cite{zeta}
 we calculate renormalized expressions
for central charge and manifest that they are free from the terms breaking
Jacobi identitites.

To begin with let us consider closed bosonic string. One can write down the 
constraints in the Hamiltonian formalism in the following form:
\begin{equation}
H_{\perp} = \frac{1}{2}p^{2} + \frac{1}{2}q'^{2},
\label{hamiltonian}
\end{equation}
\begin{equation}
H = p q',
\label{momentum}
\end{equation}
where, $H_{\perp}$ is the so called super-Hamiltonian constraint, 
$H$ is supermomentum, $q$ is the coordinate of string, while $p$ is 
conjugate momentum (we have omitted the indices enumerating  the
coordinates of d-dimensional spacetime, because the contributions of these
coordinates are addittive).
Now let us expand $q$ and its conjugate momentum $p$ via creation and
annihilation
operators and zero-mode harmonics
\begin{equation}
q = \frac{q_{0}}{\sqrt{2\pi}} + \sum_{k>0}
\frac{1}{\sqrt{2\omega_{k}}\sqrt{2\pi}}(a_{k}e^{-ikx}+\bar{a}_{k}e^{ikx}
+a_{k}^{+}e^{ikx}+\bar{a}_{k}^{+}e^{-ikx}),
\label{coord}
\end{equation}
\begin{equation}
p = \frac{p_{0}}{2\pi} + i\sum_{k>0}\sqrt{\frac{\omega_{k}}{4\pi}}
 (-a_{k}e^{-ikx}-\bar{a}_{k}e^{ikx}
+a_{k}^{+}e^{ikx}+\bar{a}_{k}^{+}e^{-ikx}),
\label{moment}
\end{equation}
where $a_{k},\bar{a}_{k},a_{k}^{+}$ and $\bar{a}_{k}^{+}$ are the
annihilation and 
creation operators for left- and right-hand oriented excitations respectively,
while $q_{0}$ and $p_{0}$ are zero modes. 

Substituting Eqs. (\ref{coord})-(\ref{moment}) into Eqs. (\ref{hamiltonian})-
(\ref{momentum}) one can write down the expressions for the harmonics of
super-Hamiltonian and supermomentum
\begin{eqnarray}
&&H_{\perp n} =
ip_{0}(\bar{a}_{n}^{+}-a_{n})\sqrt{\frac{\omega_{k}}{2}}\nonumber \\
&&-\frac{1}{2}\sum_{0<k<n}(a_{k}a_{n-k}+\bar{a}_{k}^{+}\bar{a}_{n-k}^{+})
\left(\frac{\sqrt{\omega_{k}\omega_{n-k}}}{2} + \frac{k(n-k)}{2 
\sqrt{\omega_{k}\omega_{n-k}}}\right)\nonumber \\
&&+\frac{1}{2}\sum_{k>0}(a_{n+k}\bar{a}_{k}+a_{k}^{+}\bar{a}_{n+k}^{+})
\left(\frac{k(n+k)}{ 
\sqrt{\omega_{k}\omega_{n+k}}}
-\sqrt{\omega_{k}\omega_{n-k}} 
\right)\nonumber \\
&&+\frac{1}{2}\sum_{k>0}(a_{n+k}a_{k}^{+}+\bar{a}_{k}^{+}\bar{a}_{n+k}^{+})
\left(\sqrt{\omega_{k}\omega_{n-k}} + \frac{k(n+k)}{ 
\sqrt{\omega_{k}\omega_{n+k}}}\right),
\label{hamiltonian1}
\end{eqnarray}
\begin{eqnarray}
&&H_{n} = -ip_{0}\frac{n}{\sqrt{2\omega_{n}}}(a_{n}+\bar{a}_{n}^{+})\nonumber \\
&&+\sum_{0<k<n}(\bar{a}_{k}^{+}\bar{a}_{n-k}^{+}
-a_{k}a_{n-k})\frac{k}{2}\sqrt{\frac{\omega_{n-k}}{\omega_{k}}}\nonumber \\
&&+\sum_{k>0}\frac{n+k}{2}\sqrt{\frac{\omega_{k}}{\omega_{n+k}}}
(a_{k}^{+}a_{n+k}+a_{k}^{+}\bar{a}_{n+k}^{+}
-\bar{a}_{k}a_{n+k}-\bar{a}_{k}\bar{a}_{n+k}^{+})\nonumber \\
&&+\sum_{k>0}\frac{k}{2}\sqrt{\frac{\omega_{n+k}}{\omega_{k}}}
(a_{n+k}\bar{a}_{k}+a_{n+k}a_{k}-\bar{a}_{n+k}^{+}\bar{a}_{k}
-\bar{a}_{n+k}^{+}a_{k}^{+}).
\label{momentum1}
\end{eqnarray}

As usual one can define Virasoro constraints as 
\begin{equation}
L_{n} = \frac{1}{2}(H_{\perp n} + H_{n}),
\label{Virasoro}
\end{equation}
\begin{equation}
\bar{L}_{n} = \frac{1}{2}(H_{\perp n} - H_{n}).
\label{Virasoro1}
\end{equation}
Traditional choice of dispersional relation between frequency $\omega_{k}$
and wave number $k$:
\begin{equation}
\omega_{k} = k
\label{traditional}
\end{equation}
provides diagonality of Hamiltonian $H_{\perp 0}$. Besides at the choice 
(\ref{traditional}) $L$ does not depend on 
$\bar{a},\bar{a}^{+}$ and $\bar{L}$ does not depend on $a,a^{+}$ respectively.
Other choices of dispersional relations or, in other terms, other choices of
creation
and annihilation operators obtained by means of Bogoliubov transformations
make Hamiltonian non-diagonal and the structure of Virasoro constraints
becomes more
involved. Nevertheless it is useful to write down the general expression for the
central extension for the commutators of Virasoro generators in this case as
well.
It has the following form:
\begin{eqnarray}
&&[L_{n},L_{-n}]_{c.e.} = \frac{1}{2}[H_{\perp n},H_{n}]_{c.e.}\nonumber \\
&&=\frac{1}{4}\sum_{0<k<n}\left(\omega_{k}(n-k)+\frac{k(n-k)^{2}}{\omega_{k}}
\right)\nonumber \\
&&+\frac{1}{4}\sum_{k>0}\left(\frac{(n+k)k}{\sqrt{\omega_{k}\omega_{n+k}}}
-\sqrt{\omega_{k}\omega_{n+k}}\right)\left(k\sqrt{\frac{\omega_{n+k}}{\omega
_{k}}}
-(n+k)\sqrt{\frac{\omega_{k}}{\omega_{n+k}}}\right).
\label{central}
\end{eqnarray}
If we make the traditional choice of creation and annihilation operators,
determined
by the dispersional relation (\ref{traditional}) then the second infinite
sum in the
expression (\ref{central}) disappear, while the finite sum gives the
well-known result:
\begin{equation}
[L_{n},L_{-n}]_{c.e.} = \frac{1}{2}\sum_{0<k<n}k(n-k) = \frac{1}{12}n(n^{2}-1).
\label{central1}
\end{equation}

However, the other choice of vacuum immediately implies that in the result
of finite
summation in the formula for central extension (\ref{central}) one has terms 
incompatible with Jacobi identities. On the other hand infinite sum in Eq.  
(\ref{central}) becomes divergent and needs an explicit regularization. It
is interesting 
to check how regularization of quantum commutators suggested in (\cite{HMS})
works for these
cases. 

First of all let us sketch briefly the alghorithm suggested in (\cite{HMS}).   
The main idea consisits in the substitution instead of commutator
\begin{equation}
[a_{n},a_{m}^{+}] = \delta_{nm}
\label{commutat}
\end{equation}
the regularized commutator
\begin{equation}
[a_{n},a_{m}^{+}] = \delta_{nm}f(n,s),
\label{commutat1}
\end{equation}
where $f(n,s)$ is a real function which satisfies the condition
\begin{equation}
f(n,0) = 1.
\label{condition}
\end{equation}
This regularization should give final results when the regulator is removed 
($s \rightarrow 0$). The regulator which we choose slightly differs 
from  that from Ref. (\cite{HMS}) and has the following form:
\begin{equation}
f^{(\alpha)}(n,s) = (n + \alpha)^{-s},
\label{regulator}
\end{equation}
where $\alpha$ is a positive number. 

Application of this regularization to the formula (\ref{central}) implies that
the infinite sum should be multiplied by factor
\begin{equation}
f^{(\alpha)}(k,s)f^{(\alpha)}(n+k,s),
\label{factor}
\end{equation}
because this infinite sum arises as a sum of products of commutators
\begin{equation}
[a_{k},a_{k}^{+}] [a_{k+n},a_{k+n}^{+}] 
\label{product}
\end{equation}
or the corresponding commutators with $\bar{a},\bar{a}^{+}$.

Now, let us choose for simplicity the vacuum and creation and annihilation
operators
defined by dispersional relation
\begin{equation}
\omega_{k} = \omega_{0},
\label{dispersion}
\end{equation}
where $\omega_{0}$ is a constant. Substituting (\ref{dispersion})  into 
(\ref{central}), and multiplying the terms in an infinite sum by factor
(\ref{factor}) one has
\begin{eqnarray}
&&[L_{n},L_{-n}]_{c.e.} = 
\frac{1}{4}\sum_{0<k<n}\left(\omega_{0}(n-k)+\frac{k(n-k)^{2}}{\omega_{0}}
\right)\nonumber \\
&&+\sum_{k>0}\frac{1}{4}\left(n\omega_{0} - \frac{n^{2}k}{\omega_{0}}
-\frac{nk^{2}}{\omega_{0}}\right)\frac{1}{(n+k+\alpha)^{s}} \frac{1}
{(k+\alpha)^{s}}.
\label{central2}
\end{eqnarray} 
It is easy to calculate the finite sum in the expression (\ref{central2})
using summation
formulae 
\begin{eqnarray}
&&\sum_{1}^{n}k = \frac{n(n+1)}{2},\nonumber \\
&&\sum_{1}^{n}k^{2} = \frac{n(n+1)(2n+1)}{6},\nonumber \\
&&\sum_{1}^{n}k^{3} = \frac{n^{2}(n+1)^{2}}{4}.
\label{sums}
\end{eqnarray}
Then one has
\begin{equation} 
\frac{1}{4}\sum_{0<k<n}\left(\omega_{0}(n-k)+\frac{k(n-k)^{2}}{\omega_{0}}
\right) = \frac{\omega_{0}n(n-1)}{8} + \frac{n^{2}(n+1)^{2}}{48\omega_{0}}.
\label{central3}
\end{equation}
This expression contains terms proportional to $n^{2}$ and $n^{4}$,
which as is well known break Jacobi identities for commutators. Now we
should calculate
an infinite sum in the expression (\ref{central2}) using $\zeta$-function
technique
 \cite{zeta}. 

Let us remind briefly the main formulae of this technique which we shall use.
The definition of Riemannian $\zeta$ - function is the following:
\begin{equation}
\zeta_{R}(s) = \sum_{k=1}^{\infty} \frac{1}{k^{s}}.
\label{zeta}
\end{equation}
This function is finite at $s > 1$ and can be analytically continued to the 
the field $s < 1$. In particular, we shall need the values
\begin{eqnarray}
&&\zeta_{R}(0) = -\frac{1}{2},\nonumber \\
&&\zeta_{R}(-1) = -\frac{1}{12},\nonumber \\
&&\zeta_{R}(-2) = 0.
\label{zeta1}
\end{eqnarray}
At $s = 1$ Riemannian $\zeta$ - function has the pole:
\begin{equation}
\zeta_{R}(1+s) = \frac{1}{s} + const + 0(s).
\label{zeta2}
\end{equation}

Now, expanding factor depending on $s$ in the infinite sum in expression 
(\ref{central2}) in inverse degrees of $k$ one can get the following expression:
\begin{eqnarray}
&&\frac{1}{(n+k+\alpha)^{s}} \frac{1}{(k+\alpha)^{s}} = k^{-2s} - 2sk^{-2s-1}
\left(\alpha + \frac{n}{2}\right)\nonumber \\
&&+s(2s+1)k^{-2s-2}\left(\alpha + \frac{n}{2}\right)^{2}-s(2s+1)(2s+2)
k^{-2s-3}\frac{\left(\alpha + \frac{n}{2}\right)^{3}}{3}\nonumber \\
&&+sk^{-2s-2}\frac{n^{2}}{4}-s(s+1)k^{-2s-3}
\left(\alpha + \frac{n}{2}\right)\frac{n^{2}}{2}.
\label{regulator1}
\end{eqnarray}
Contracting expression (\ref{regulator1}) with the multiplier
$\left(n\omega_{0} - \frac{n^{2}k}{\omega_{0}}
-\frac{nk^{2}}{\omega_{0}}\right)$ in the expression (\ref{central2}) one
can get the 
limit $s \rightarrow 0$ using formulae (\ref{zeta1}), (\ref{zeta2}). As a
result one
can get renormalized contribution of the infinite sum in the expression 
(\ref{central2})
\begin{eqnarray}
&&\left\{\sum_{k>0}\frac{1}{4}\left(n\omega_{0} - \frac{n^{2}k}{\omega_{0}}
-\frac{nk^{2}}{\omega_{0}}\right)\right\}_{renormalized}\nonumber \\
&&=
n\left(\frac{\omega_{0}(1-2\alpha)}{8}+\frac{\alpha^{3}}{12\omega_{0}}\right)
\nonumber \\
&&+n^{2}\left(\frac{1}{48\omega_{0}} - \frac{\omega_{0}}{8}\right)
-\frac{n^{4}}{48\omega_{0}}.
\label{renorm}
\end{eqnarray}
It is easy to see that dangerous terms proportional to $n^{2}$ and $n^{4}$  
in the regularized expression for the infinite sum (\ref{renorm}) exactly cancel
those in the the expression for the finite sum (\ref{central3}). Thus in such a 
way ultraviolet renormalization given by regulator (\ref{commutat1}), 
(\ref{condition}) and (\ref{regulator}) provides the restoration of Jacobi
identities.

One can consider also another choice of creation and annihilation operators 
defined by the dispersional relation 
\begin{equation}
\omega_{k} = A k^{2}.
\label{dispersion1}
\end{equation}
One can easily check that all the results of calculations for the case given by
dispersion relation (\ref{dispersion1}) coincide with those obtained in the
case 
given by dispersion relation (\ref{dispersion}).
(All the formulae for the latter case can be obtained from the formulae for
the  
former case by substitution $A \rightarrow 1/\omega_{0}$).

Thus, we have seen that the regularization of commutators suggested in Ref. 
(\cite{HMS}) applied to the string model with non-standard definitions of
creation and annihilation operators allows to preserve fundamental symmetric
properties encoded in Jacobi identities. This could be interpreted as an 
additional argument in favour of this regularization. However, we should
recognize
that it is not true for any choice of vacuum. For example, for the vacuum
defined
by dispersion relation
$$
\omega_{k} = \frac{B}{k}
$$
(for this case all the calculations also could be carried out in an explicit
form)
we have got the term proportional to $n^{5}$, which cannot be cancelled by
help of
regularization of creation and annihilation operators. The analysis of
applicability
of this regularization to different types of vacuums and its application to 
such complicated theories as quantum cosmology is under investigation now
\cite{future}.

A.K. is grateful to CARIPLO Science Foundation for financial support.
Work of A.K. and I.K. was partially supported by RFBR via grants 96-02-16220
and 
96-15-96458.

\end{document}